\documentstyle[aps,psfig]{revtex}

\begin{document}

\draft
%\preprint{}
\title{
Measurement of the flux and zenith-angle distribution of upward
through-going muons in Kamiokande II+III
}
\vspace{0.3cm}
\author{
S. Hatakeyama$^1$, T. Hara$^2$, Y. Fukuda$^3$, T. Hayakawa$^3$, K.
Inoue$^3$, K. Ishihara$^3$, H. Ishino$^3$,
S. Joukou$^3$, T. Kajita$^3$, S. Kasuga$^3$, Y. Koshio$^3$,
T. Kumita$^{3}$\cite{kumikumi},
K. Matsumoto$^3$, M. Nakahata$^3$, K. Nakamura$^3$\cite{byline}, 
K. Okumura$^3$,
A. Sakai$^3$\cite{byline}, M. Shiozawa$^3$, J. Suzuki$^3$,
Y. Suzuki$^3$,
T. Tomoeda$^3$, Y. Totsuka$^3$,
K. S. Hirata$^4$, K. Kihara$^4$, Y. Oyama$^4$\cite{byline},\\
M. Koshiba$^5$\cite{diamond},
K. Nishijima$^6$,
T. Horiuchi$^6$,
K. Fujita$^1$, M. Koga$^1$, T. Maruyama$^1$,
A. Suzuki$^1$, M. Mori$^7$\cite{taketake},\\
T. Suda$^8$\cite{sudasuda},
A. T. Suzuki$^8$,
T. Ishizuka$^9$, K. Miyano$^9$, H. Okazawa$^9$,
Y. Nagashima$^2$, M. Takita$^2$,
T. Yamaguchi$^2$,\\
Y. Hayato$^a$\cite{byline},
K. Kaneyuki$^a$, T. Suzuki$^a$
Y. Takeuchi$^a$\cite{taketake}, T. Tanimori$^a$,
S. Tasaka$^b$, E. Ichihara$^c$,\\
S. Miyamoto$^c$,
and K. Nishikawa$^c$\cite{byline}
}
\vspace{0.5cm}
\address{
$^1$Physics Department, Graduate School of Science, Tohoku University,
Sendai, Miyagi 980-8578, Japan
}
\address{
$^2$High Energy Physics Division, Graduate School of Science,
Osaka University,
Toyonaka, Osaka 560-0043, Japan
}
\address{
$^3$Institute for Cosmic Ray Research, University of Tokyo,
Tanashi, Tokyo 188-8502, Japan
}
\address{
$^4$National Laboratory for High Energy Physics,
Tsukuba, Ibaraki 305-0801, Japan
}
\address{
$^5$Institute of Research and Development, Tokai University,
Shibuya, Tokyo 151-0063, Japan
}
\address{
$^6$Department of Physics, Tokai University, Hiratsuka,
Kanagawa 259-1292, Japan
}
\address{
$^7$Department of Physics, Miyagi University of Education,
Sendai, Miyagi 980-0845, Japan
}
\address{
$^8$Department of Physics, Kobe University,
Kobe, Hyogo 657-8501, Japan
}
\address{
$^9$Niigata University, Niigata 950-2181, Japan
}
\address{
$^a$Department of Physics, Tokyo Institute of Technology,
Megro-ku, Tokyo 152-8551, Japan
}
\address{
$^b$Department of Physics, Gifu University,
Gifu, Gifu 501-1193, Japan
}
\address{
$^c$Institute for Nuclear Study, University of Tokyo,
Tanashi, Tokyo 188-8501, Japan
}
\date{\today}
\maketitle
\begin{abstract}
The flux of upward through-going
muons of minimum (mean) threshold energy $>$1.6 (3.0) GeV is measured,
based on a total of 372 events observed
by the Kamiokande~II+III detector during 2456 detector live days.
The observed muon flux was
$\Phi_{obs}=(1.94\pm0.10  {\mbox{(stat.)}}^{+0.07}_{-0.06}
{\mbox{(sys.)}})\times10^{-13}
{\rm{cm^{-2}s^{-1}sr^{-1}}}$, which is compared to an expected value
of $\Phi_{theo}=(2.46\pm0.54(\rm{theo.}))\times
10^{-13}{\rm{cm^{-2}s^{-1}sr^{-1}}}$.
The observation is in agreement with the prediction within the errors.
The zenith angle dependence of the
observed upward through-going muons 
supports the previous indication of neutrino oscillations
made by Kamiokande using sub- and multi-GeV atmospheric neutrino
events. 
\end{abstract}
%\pacs{PACS numbers: 14.60.Pq, 96.40.Tv}

\vspace{0.5cm}

%\narrowtext

In 1988, the Kamiokande group observed a smaller
($\nu_\mu+\overline\nu_\mu$)/($\nu_e+\overline\nu_e$) ratio
than the expected value in sub-GeV fully-contained
atmospheric neutrino events
\cite{hirata1,takita}.
This discrepancy was later confirmed by other experiments
IMB-3 \cite{casper,szendy} and Soudan II \cite{allison},
while some other experiments
(Fr\'{e}jus \cite{berger} and Nusex \cite{aglietta})
did not observe such discrepancy within the experimental errors.

Recently, a measurement of this ratio using multi-GeV 
atmospheric neutrino events was made by the Kamiokande 
experiment\cite{fukuda}.
The observed
($\nu_\mu+\overline\nu_\mu$)/($\nu_e+\overline\nu_e$) ratio in this
energy range was also 
significantly smaller than expected and depended strongly on the
zenith angle.

Having observed the discrepancy in both sub- and
multi-GeV atmospheric neutrino events, 
the next logical step \cite{oyama,becker,boliev,macro} is to investigate
events that are produced by neutrinos
of even higher energies.
Energetic atmospheric $\nu_{\mu}$ or $\overline{\nu}_{\mu}$ traveling
through  Earth
interacts with rock layers surrounding the detector and produce muons
via charged-current interactions.
Though neutrino-induced downward-going muons hit the detector,
these downward-going muons are
difficult to differentiate from cosmic-ray muons.
On the contrary, upward-going muons are considered
neutrino-induced, since upward-going cosmic-ray muons
range out in Earth.
Especially, those energetic enough to pass through the detector
are defined as ``upward through-going muons''.
The mean energy of the parent neutrino producing them is approximately
100 GeV.

The experimental site is located 1000 m underground
in the Kamioka mine, Gifu prefecture, Japan.
The Kamiokande-II detector is a cylindrical water
Cherenkov calorimeter.
The inner detector is 14.4 m in diameter $\times$ 13.1 m in height,
which contains 2142 metric tons of purified water.
A total of 948 20-inch-in-diameter photomultiplier tubes (PMTs) are
mounted on the tank walls and covers 20\% of the inner tank surface.
The detector also has an optically isolated $4\pi$ solid-angle
anticounter which is also a 1500-metric-ton
water Cherenkov detector with 123 
20-inch-in-diameter PMTs and is divided into top, barrel and bottom
parts.
The anticounter tags incoming and/or outgoing muons and
to shield $\gamma$-rays and neutrons from the surrounding rock. 
Charge and relative timing information from each hit PMT is recorded 
for the event reconstruction.
The triggering efficiency for a muon having momenta more than 200
MeV/c is $\sim$100\%.
The nominal detector effective area for upward-going muons
is roughly 150 ${\rm{m}^2}$.
The deviation between the reconstructed track direction and
the real muon direction ($\Delta\theta_{rec}$) is studied by using
Monte Carlo events and is
estimated to be $2.1^\circ$ \cite{oyama}.
See \cite{takita,oyama}  for more details of the 
Kamiokande-II detector. The new phase of the experiment started in
December 1990, after replacement of the entire electronics,
replacement of the dead PMTs and installation of an aluminized mylar
cone around each PMT.
The description of the Kamiokande-III detector is found
elsewhere~\cite{hara}. 

The data sets used in this analysis are taken from December 1985 to 
April 1990 in Kamiokande-II and from December 1990 to
May 1995 in Kamiokande-III corresponding to 2456-day detector
livetime. A total of $2.2\times10^8$ events were recorded during these
data-taking periods.

In Kamiokande-III, when the following three requirements are
satisfied, it is considered 
that a muon hits one of the three anticounter parts: (1) total number of
photoelectrons (p.e.) in a part of the anticounter $\geq$ 20 p.e.
(2) number of hit PMT in the same part of the anticounter $\geq$ 5
(3) Maximum number of photoelectrons/PMT in an event in the same
part of the anticounter $\geq$ 5 p.e.
By requiring the two parts or the barrel part hit in the anticounter, 
through-going muons can be selected.
A selection criterion that the total number of photoelectrons of an
event should be larger than 6000 p.e. is imposed to ensure the minimum
muon track length of 7~m (corresponding to minimum (mean) muon
threshold energy of 1.6 (3.0) GeV).
An upper limit is also set at 40000 p.e. (corresponding to $\simeq$
30~m track length for a relativistic muon in the detector) to remove
multiple muons or
muons accompanied with bremsstrahlung, pair production and/or hadronic
showers. The event reduction procedures in Kamiokande-II is described
elsewhere~\cite{oyama}.

To eliminate abundant downward-going cosmic-ray muons (0.37 Hz), 
the events satisfying $\cos\Theta>-0.04$ are cut.
where $\Theta$ is the zenith angle with $\cos\Theta=1$ corresponding
to downward-going events.
The detection efficiency for upward through-going muons is estimated
by a Monte Carlo simulation.
Using the upward/downward symmetry of the detector
configuration, the validity of the Monte Carlo
program is checked by  real cosmic-ray downward
through-going muons.
The average detection efficiency is estimated to be $\sim97\%$.
The details of the selection criteria are described in \cite{hara}.
Finally, 184 and 188 upward through-going muon events are
observed during Kamiokande-II \cite{oyama,mori} (1124 days) and
Kamiokande-III \cite{hara}
(1332 days) experimental periods, respectively.

In this analysis, the combination of the Bartol 
atmospheric neutrino flux model \cite{gaisser}, 
the parton distribution functions of GRV94DIS \cite{grv94} and the
Lohmann's muon energy loss formula in the standard rock \cite{lohmann}
are employed, when one analytically calculates the standard expected
muon flux.
Also a Monte Carlo technique is introduced in order to
estimate the directionality between a muon and the parent neutrino 
and it is found that
$\sqrt{\Delta\theta_{\nu\mu}^2+\Delta\theta_{mul}^2}$
is $4.1^\circ$ for the atmospheric neutrino energy
spectrum, where 
$\Delta\theta_{\nu\mu}$ stands for 
$\mu$ production angle relative to $\nu$ direction in the
neutrino-nucleon interaction
and $\Delta\theta_{mul}$ for deflection by
multiple Coulomb scatterings in the rock.
To estimate the model-dependent uncertainties ($\pm$10 \% for the
absolute flux normalization and $-$3.6\% to +1.5 \% for the bin-by-bin 
shape difference
in zenith-angle distribution) of the
expected muon flux,
the combination of another atmospheric neutrino
flux model calculated by Honda et al. \cite{honda}
and another
parton distribution functions of CTEQ3M \cite{cteq} are
also considered.  

As a result,  the expected muon
flux $\Phi_{theo}$is calculated to be
$(2.46\pm0.54(\rm{theo.}))\times10^{-13}{\rm{cm^{-2}}}$
${\rm{s^{-1}sr^{-1}}}$ (cos$\Theta$\(<\)$-$0.04),
where the estimated theoretical uncertainties are 
described in Table \ref{systematictable}. The dominant
one originates from the absolute normalization uncertainty
in the neutrino flux which is estimated to be approximately $\pm20$
\%\cite{gaisser,honda,frati} above several GeV.

Given the
detector live time, $T$, the effective area for upward through-going
muons, $S(\Theta)$, and the detection efficiency,
$\varepsilon(\Theta)$, 
the upward through-going muon flux is calculated by the
formula:

\[
\Phi_{obs}=\sum^{N}_{j=1}\left(
\frac{1}{\varepsilon_2(\Theta_j)T_2+\varepsilon_3(\Theta_j)T_3}\right)
\cdot\frac{1}{S(\Theta_j)2(1-0.04)\pi}
\]

\noindent
where suffix 2 (3) represents Kamiokande-II (-III),
the suffix $j$ stands for each event number,
$2(1-0.04)\pi$ is
the total solid angle covered by the detector for upward through-going
muons,
$N$ corresponds to the total number of observed
muon events (372 events).
The angular dependence of the detection efficiency 
for each experimental period
is found in \cite{oyama} and \cite{hara}, respectively.
Conceivable experimental systematic errors are summarized in
Table~\ref{systematictable}.

Taking these experimental systematic errors into account, the observed 
upward through-going muon flux is:

\[
\Phi_{obs}=(1.94\pm0.10 {\mbox{(stat.)}}^{+0.07}_{-0.06}
{\mbox{(sys.)}}) 
\times10^{-13}{\rm{cm^{-2}s^{-1}sr^{-1}}}.
\]

At present, three experiments other than Kamiokande reported
the measurement of the upward through-going muon flux and
their results are summarized in Table~\ref{uptable}, together
with this measurement.
The observed flux by each experiment is 
in agreement with the expectation within
the errors.

\begin{figure}[htbp]
\centering
\leavevmode
\psfig{file=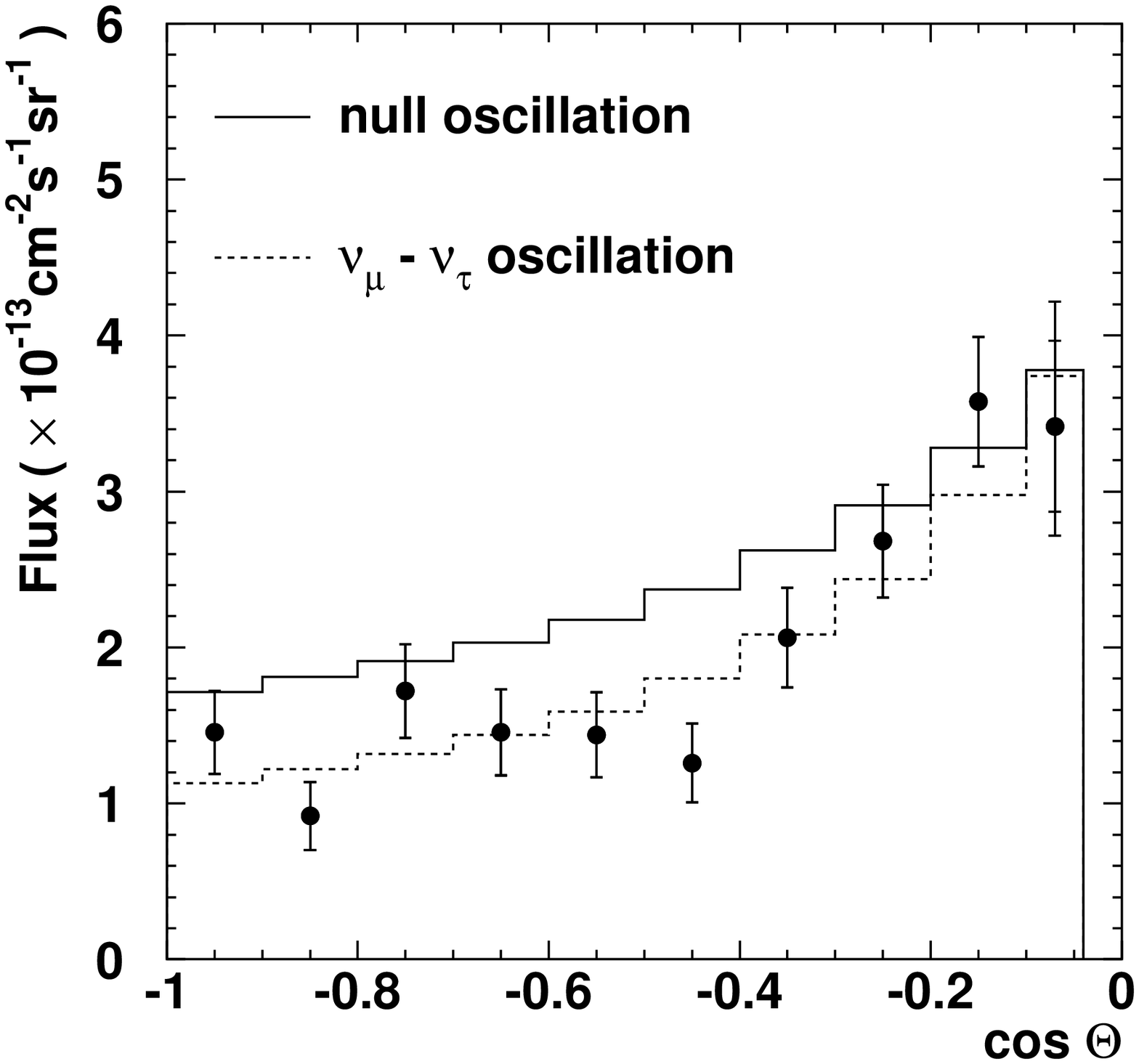,height=17cm}
\caption{Zenith angle distribution of upward through-going muon
flux observed in Kamiokande II+III. Inner (outer) error bars indicate 
statistical (uncorrelated experimental systematic + statistical added
in quadrature) errors. 
The solid histograms show the expected (Bartol + GRV94DIS + Lohmann's
ionization formula) upward through-going muon flux
for the null neutrino oscillation case. Also shown is the expected
flux (dashed) assuming the
best fit parameters at $(\sin^2 2\theta, \Delta m^2)$=$(1.00,
3.2\times10^{-3}{\rm{eV}}^{2})$,
$\alpha$=1.00 in the physical region
for 
the $\nu_{\mu}\leftrightarrow\nu_{\tau}$ oscillation case  
.}
\label{k23angdist}
\end{figure}

The zenith angle distribution of the observed flux 
$(d\Phi/d\Omega)_{obs}^{i}$ is shown
in Fig.~\ref{k23angdist}.
The shape of the distribution is not well represented by the
theory ($\chi^{2}$/degrees of freedom = 21.3/9.)

A neutrino oscillation hypothesis is then tested using the
zenith angle distribution.
The expected flux for a
given set of $\Delta m^{2}$ and $\sin^{2}2\theta$ is calculated 
and the same binning is applied as data.

To test the validity of the oscillation hypothesis,
a $\chi^{2}$ is calculated as 

\[
\chi^{2}={\min}\left[
\sum_{i=1}^{10}\left(\frac
{\left(\frac{d\Phi}{d\Omega}\right)_{obs}^{i}-
\alpha\left(\frac{d\Phi}{d\Omega}\right)_{osc}^{i}}
{\sqrt{\sigma_{stat,i}^{2}+\sigma_{sys,i}^{2}}}\right)^{2}+
\left(\frac{\alpha-1}{\sigma_{\alpha}}\right)^{2}\right]
\]

\noindent
where $\sigma_{stat,i}$  ($\sigma_{sys,i}$) is the statistical
(experimental systematic)  error in
$(d\Phi/d\Omega)_{obs}^{i}$ 
in the $i$-th bin, 
$\alpha$ is an absolute normalization factor of the expected flux.
Based on uncorrelated systematic errors in
Table~\ref{systematictable} added in quadrature, we estimate 
$\sigma_{sys,i}$ to be $\pm$18.5 \% for $-$0.1\(<\)cos$\Theta$\(<\)$-$0.04
and $\pm$1.2 to $\pm$3.8 \% for $-1\le$cos$\Theta\le -0.1$, respectively.  
Although the nature of the uncertainty in $\alpha$ is
unknown, it is presumed that $\alpha$ obeys a Gaussian distribution
with one standard deviation error $\sigma_\alpha$.
The absolute flux normalization error $\sigma_\alpha$
is estimated to be
$\pm$22~\% by adding in quadrature the correlated experimental 
errors and 
theoretical uncertainties
in Table~\ref{systematictable}.
Then, the minimium $\chi^{2} (\chi^{2}_{min})$ on the $\Delta
m^{2}-\sin^{2}2\theta$ plane is searched for.

\begin{figure}[htbp]
\centering
\leavevmode
\psfig{file=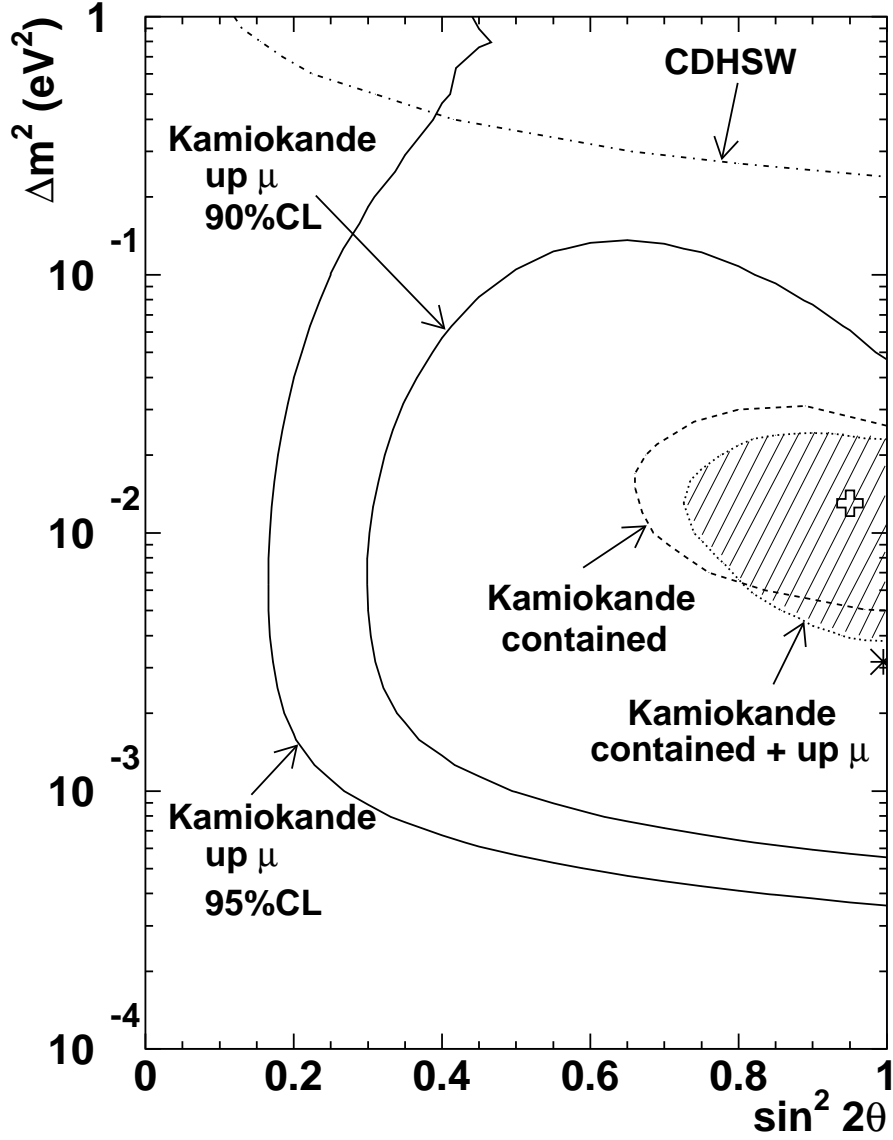,height=17cm}
\caption{The allowed regions at 90~\% C.L., obtained
from
the Kamiokande II+III upward through-going muon analysis(up $\mu$) and
from the combined analysis (Kamiokande
contained\protect{\cite{fukuda}} + up $\mu$), drawn on the
($\sin^{2}2\theta$,$\Delta{m}^{2}$) plane for
$\nu_{\mu}\leftrightarrow\nu_{\tau}$ oscillations.
Also shown are the excluded region by CDHSW\protect{\cite{cdhsw}} and
the allowed region by    
the Kamiokande contained event analysis\protect{\cite{fukuda}}. The 
right-hand side of the contours are allowed (excluded) regions by
Kamiokande (CDHSW). 
The asterisk (cross) indicates
the best fit point 
for up $\mu$  (for Kamiokande contained + up $\mu$) at $(\sin^2
2\theta, \Delta m^2)=(1.00, 3.2\times 10^{-3}{\rm{eV}^2})$
(at $(\sin^2 2\theta, \Delta m^2)=(0.95, 1.3\times
10^{-2}{\rm{eV}^2})$).
The 95~\% C.L. contour (up $\mu$) is presented as well.} 
\label{nmntcontour90}
\end{figure}

We note here that the allowed region 
given by Kamiokande sub- and
multi-GeV contained event analysis~\cite{fukuda} 
for $\nu_{e}\leftrightarrow\nu_{\mu}$ oscillations
has been already excluded by the CHOOZ experiment~\cite{chooz},
suggesting $\nu_{\mu}\leftrightarrow\nu_{\tau}$ oscillations
for the contained and upward-going muon events. 
Assuming $\nu_{\mu}\leftrightarrow\nu_{\tau}$ oscillations,
$\chi^{2}_{min}(=12.2)$ 
occurs  
in the unphysical region at
 $(\sin^2 2\theta, \Delta m^2)=
(1.35, 2.0\times 10^{-3}{\rm{eV}^2})$ and $\alpha$=1.02.
On the other hand, if we bound the oscillation parameters
in the physical region, $\chi^{2}_{min}$(=12.8)
takes place at  
 $(\sin^2 2\theta, \Delta m^2)=
(1.00, 3.2\times 10^{-3}{\rm{eV}^2})$ and $\alpha$=1.00.
For the null oscillation case, we obtain $\chi^{2}_{min}$
of 21.3 at $\alpha$=0.77 (probability of statistical fluctuation: 
1\%).  
The zenith angle distribution of
$\alpha(d\Phi/d\Omega)^{i}_{osc}$ for the best fit parameters
in the physical region
is shown in Fig.~\ref{k23angdist} together with
the data.  
Figure~\ref{nmntcontour90} shows the
allowed region contours on the $(\sin^2 2\theta, \Delta m^2)$
plane for
$\nu_{\mu}\leftrightarrow\nu_{\tau}$ oscillations.
As $\chi^{2}_{min}$ for $\nu_{\mu}\leftrightarrow\nu_{\tau}$
oscillations falls down in the unphysical region, 
the contours are drawn according to the 
prescription for bounded physical regions given in Ref.\cite{pdg}.
If we replace the Bartol neutrino flux\cite{gaisser} by the 
Honda's\cite{honda} and/or
the GRV94DIS parton distribution functions\cite{grv94} 
by the CTEQ3M's\cite{cteq}, the allowed region
contours are similar to those presented in Fig.~\ref{nmntcontour90}.
Consequently, we find that the zenith angle dependence is in favor of
the $\nu_{\mu}\leftrightarrow\nu_{\tau}$ oscillation hypothesis 
and supports the Kamiokande 
sub- and multi-GeV contained event analysis~\cite{fukuda}.

Combining this work with the previous analysis~\cite{fukuda} of the
Kamiokande 
sub- and multi-GeV contained
events, we draw an allowed region contour 
on the $(\sin^2 2\theta, \Delta m^2)$
plane, as is shown 
in Fig.~\ref{nmntcontour90}.
The best fit point $(\sin^2 2\theta, \Delta m^2)=
(0.95, 1.3\times 10^{-2}{\rm{eV}^2})$ is obtained
for 
$\nu_{\mu}\leftrightarrow\nu_{\tau}$ oscillations.

%\narrowtext

In conclusion, based on 372 upward through-going muon events during
2456 detector live days,
the flux of the upward through-going muons ($>$1.6 GeV) produced
by atmospheric muon neutrinos in the rock layers surrounding the
detector is
measured with the Kamiokande~II+III detector:
$\Phi_{obs}=(1.94\pm0.10 {\mbox{(stat.)}}^{+0.07}_{-0.06}
{\mbox{(sys.)}})\times10^{-13}
{\rm{cm}^{-2}{s}^{-1}{sr}^{-1}}$.
This is compared with the expected flux
calculation of
$\Phi_{theo}=(2.46\pm0.54(\rm{theo}.))\times10^{-13}
{\rm{cm}^{-2}{s}^{-1}{sr}^{-1}}$. 
The observed upward through-going muon flux is in agreement
with the expected flux within the relatively large uncertainties in
the theoretical calculations.
We find that the zenith angle dependence of the upward through-going
muons does not agree with the theoretical expectation 
(probability of statistical fluctuation: 1\%), yet is favored by 
the $\nu_{\mu}\leftrightarrow\nu_{\tau}$ oscillation assumption.
This result supports the indication of neutrino oscillations given
by the analysis of the sub- and multi-GeV atmospheric 
neutrino events by Kamiokande.

We gratefully acknowledge the cooperation of the Kamioka Mining and 
Smelting Company.
This work was supported by the Japanese Ministry of Education,
Science, Sports and Culture.

\newpage

\newpage

\newpage
\begin{table}[htbp]
\centering
\begin{tabular}{c|c}
Error source & Error (\%) \\
\hline
uncertainty in $\Delta\theta_{rec}$ & $\pm$1$^{\rm{a}}$\\
cos$\Theta=-0.04$ cut dependence & +18.2, $-4.4^{\rm{a}}$\\
cosmic ray $\mu$ contamination & $-10^{\rm{a}}$ \\
detection efficiency & +2.8, $-8.3^{\rm{a}}$  \\
 & +0.6, $-1.2^{\rm{b}}$\\
7~m track length cut & $\pm$0.5$^{\rm{c}}$ \\
live time & $\pm$0.4$^{\rm{c}}$ \\
effective area & $\pm$0.3$^{\rm{c}}$ \\
PMT gain & $\ll$1$^{\rm{c}}$ \\
water transparency & $\ll$1$^{\rm{c}}$ \\
chemical component in the rock &  $\ll$1$^{\rm{d}}$ \\ 
$\nu$ flux normalization\%\cite{gaisser,honda,frati} & $\pm$20$^{\rm{d}}$ \\
theoretical model dependence & $\pm$10$^{\rm{d}}$ \\
 & $-3.6$ to +1.5$^{\rm{e}}$
%\hline
%total & $\pm$23 \\
 
\end{tabular}
\caption{List of experimental systematic errors and theoretical 
uncertainties in the flux measurement.
a: experimental uncorrelated systematic error specific in the most
horizontal bin $-$0.1\(<\)cos$\Theta$\(<\)$-$0.04,
b: experimental uncorrelated systematic error specific in the bins
$-1\le$cos$\Theta\le-0.1$,
c: experimental correlated systematic error,
d: theoretical correlated uncertainty,
e: theoretical uncorrelated uncertainty. }
\label{systematictable}
\end{table}

\newpage
\begin{table}[htbp]
%\widetext
\centering
\begin{tabular}{c|ccc}
Experiment & Minimum E$_{\mu}\rm{(GeV)}$ & Observed & Expected \\
\hline
${\rm{Kamiokande}}$ & 1.6
& $\begin{array}{c}
\vspace{-1mm}(1.94\pm0.10(\rm{stat.})^{+0.07}_{-0.06}(\rm{sys.}))\\ 
\times10^{-13}{\rm{cm^{-2}s^{-1}sr^{-1}}}
  \end{array} $
& $\begin{array}{c}
\vspace{-1mm}(2.46\pm0.54(\rm{theo.})) \\
\times10^{-13}{\rm{cm^{-2}s^{-1}sr^{-1}}}
  \end{array}$ \\
${\rm{IMB}}$\cite{becker} & $\sim$1.4 & 532$\pm 23(\rm{stat.})$ events
& 516$\pm 103(\rm{theo.})$ events \\
${\rm{Baksan}}$\cite{boliev} & 1.0 &
559 events &
580 events \\
${\rm{MACRO}}$\cite{macro} & 1.0 &
$277\pm17(\mbox{stat.})\pm22(\mbox{sys.})$events &
$371\pm63(\mbox{theo.})$events \\
\end{tabular}
\caption{Summary of upward-going muon flux measurement.  E$_{\mu}$
represents the muon energy. Kamiokande, Baksan and Macro employ
the Bartol neutrino flux\protect{\cite{gaisser}}, while IMB adopts the
Lee and Koh neutrino flux\protect{\cite{lee}} 
(the Volkova flux\protect{\cite{volkova}}) for neutrino 
energies lower (higher) than 15 GeV.}
\label{uptable}
%\narrowtext
\end{table}


\begin{references}
  \bibitem[\dag]{kumikumi}Present address: Department of Physics,
Tokyo Metropolitan University, Hachiouji, Tokyo 192-0397, Japan.
  \bibitem[*]{byline}Present address: High Energy Accelerator Research
Organization (KEK), Tsukuba, Ibaraki 305-0801, Japan.
  \bibitem[\diamond]{diamond}Present address: The University of Tokyo,
Tokyo 113-0033, Japan.
  \bibitem[\S]{sudasuda}Deceased.
  \bibitem[\P]{taketake}Present address: Institute for Cosmic Ray
Research, University of Tokyo, Tanashi, Tokyo 188-8502, Japan.
  \bibitem{hirata1}K. S. Hirata $et$ $al.$, Phys.\ Lett.\ {\bf B205},
416 (1988); K. S. Hirata $et$ $al.$, Phys.\ Lett. \ {\bf B280}, 146
(1992). 
  \bibitem{takita}M. Takita, Ph. D. thesis, Fac. of Science,
Univ. of Tokyo(1989), ICR-Report-186-89-3.
  \bibitem{casper}D. Casper $et$ $al.$, Phys.\ Rev.\ Lett.\ {\bf 66},
2561 (1991). 
  \bibitem{szendy}R. Becker-Szendy $et$ $al.$, Phys.\ Rev.\ {\bf D46},
3720 (1992).
  \bibitem{allison}W. W. Allison $et$ $al.$, Phys. Lett.  {\bf B391},
491 (1997).
  \bibitem{berger}Ch. Berger $et$ $al.$, Phys. Lett. {\bf B227}, 489
(1989); K. Daum $et$ $al.$, Z. Phys. C {\bf 66}, 417 (1995).
  \bibitem{aglietta}N. Aglietta $et$ $al.$, Europhys.\ Lett. {\bf 8},
611 (1989). 
  \bibitem{fukuda}Y. Fukuda $et$ $al.$, Phys.\ Lett.\ {\bf B335}, 237
(1994). 
  \bibitem{oyama}Y.Oyama $et$ $al.$, Phys.\ Rev. \ {\bf D39 }, 1481
(1989); Y. Oyama, Ph. D. thesis, Fac. of Science, Univ. of
Tokyo(1989), ICR-Report-193-89-10.
  \bibitem{becker}R. Becker-Szendy $et$ $al$., Phys.\ Rev.\ Lett.
\ {\bf 69}, 1010 (1992).
  \bibitem{boliev}M. M. Boliev et al., Proc. of the 24$^{\rm{th}}$ 
ICRC vol. 1, p722 (1995); T. K. Gaisser, Proc. of the 17$^{\rm{th}}$
Int. Conf. on Neutrino Physics and Astrophysics NEUTRINO'96, p211
(1997). 
  \bibitem{macro}P. Bernardini (MACRO Collaboration), Proc. of the
25$^{\rm{th}}$ ICRC 4.6.44 (1997); S. Ahlen  $et$ $al.$, Phys.\ Lett.\
{\bf B357}, 481 (1995).
  \bibitem{hara}T. Hara, Ph. D. thesis, Osaka University(1996); S.
Hatakeyama, Ph. D. thesis, Tohoku University(1998).
  \bibitem{mori}M. Mori et al., Phys.\ Lett.\ {\bf B270}, 89 (1991).
  \bibitem{gaisser}V. Agrawal, T. K. Gaisser, P. Lipari, T. Stanev,
Phys.\ Rev.\ {\bf D53}, 1314 (1996).
  \bibitem{grv94}M. Gl\H{u}ck, E. Reya and A. Vogt,
Z.\ Phys.\ {\bf C67}, 433 (1995).
  \bibitem{lohmann}W. Lohmann, R. Kopp and R. Voss, CERN Yellow Report
No. 85-03.
  \bibitem{honda}M. Honda $et$ $al.$, Phys.~Rev.~{\bf D52}, 4985
(1995), Prog.\ Theor.\ Phys.\ Suppl.\ {\bf 123}, 483 (1996).
  \bibitem{cteq}J. Botts $et$ $al.$, Phys.\ Lett.\ {\bf B304}, 159
(1993); Lai H. L. $et$ $al.$, Phys.\ Rev. {\bf D51}, 4763 (1995).
  \bibitem{frati}W. Frati $et$ $al.$, 
Phys.\ Rev. {\bf D48}, 1140 (1993).
\bibitem{lee}H. Lee and Y. S. Koh, Nuovo Cimento {\bf B105}, 883
(1990). 
\bibitem{volkova}L. V. Volkova, Yad. Fiz. {\bf 31}, 1510 (1980) (Sov.
J. Nucl. Phys. {\bf 31 }, 784 (1980)).
\bibitem{chooz}M. Apollonio $et$ $al.$, hep-ex/9711002, 1997.
\bibitem{pdg}Particle properties data booklet, 
Section ^^ ^^ Gaussian errors - bounded physical region'', June
(1992). 
 \bibitem{cdhsw}F. Dydak $et$ $al.$, Phys.\ Lett.\ {\bf B134},
281 (1984).
\end{references}
\end{document}